\documentclass[fleqn,10pt]{wlscirep}
\usepackage{amssymb}
\usepackage{graphicx}
\usepackage{dcolumn}
\usepackage{bm}
\usepackage{color}

\begin{document}

\newcommand{\ba}{{\bf a}}
\newcommand{\bbb}{{\bf b}}
\newcommand{\br}{{\bf r}}
\newcommand{\bp}{{\bf p}}
\newcommand{\bq}{{\bf q}}
\newcommand{\bk}{{\bf k}}
\newcommand{\bg}{{\bf g}}
\newcommand{\bs}{{\bf s}}
\newcommand{\bt}{{\bf t}}
\newcommand{\bG}{{\bf G}}
\newcommand{\bP}{{\bf P}}
\newcommand{\bJ}{{\bf J}}
\newcommand{\bK}{{\bf K}}
\newcommand{\bL}{{\bf L}}
\newcommand{\bR}{{\bf R}}
\newcommand{\bS}{{\bf S}}
\newcommand{\bT}{{\bf T}}
\newcommand{\bA}{{\bf A}}
\newcommand{\bH}{{\bf H}}

\newcommand{\ga}{\alpha}
\newcommand{\gm}{\mu}
\newcommand{\gb}{\beta}
\renewcommand{\gg}{\gamma}
\newcommand{\gd}{\theta}
\newcommand{\ep}{\epsilon}
\newcommand{\gl}{\lambda}
\newcommand{\go}{\omega}


\title{Tuning topological surface magnetism by bulk alloying}

\author[1]{N. Klier}
\author[2]{S. Sharma}
\author[1,*]{O. Pankratov}
\author[1,*]{S. Shallcross}
\affil[1]{Lehrstuhl f\"ur Theoretische Festk\"orperphysik, Staudtstr. 7-B2, 91058 Erlangen, Germany,}
\affil[2]{Max-Planck-Institut f\"ur Mikrostrukturphysik Weinberg 2, 06120 Halle, Germany.}
\affil[*]{oleg.pankratov@fau.de}
\affil[*]{sam.shallcross@fau.de}
\date{\today}

\begin{abstract}

Deploying an analytical atomistic model of the bulk band structure of the IV-VI semiconductors we connect the spin structure of the topological surface state to the crystal field and spin orbit coupling parameters of the bulk material. While the Dirac-Weyl (or equivalently, Rashba) type topological surface state is often assumed universal, we show that the physics of the surface state is strikingly non-universal. To see this explicitly we calculate the RKKY interaction, which may be viewed as a probe of this surface state spin structure, finding its \emph{qualitative form} depends on the values the bulk spin-orbit and crystal field parameters take. This opens the way to tune the spin interaction on the surface of a IV-VI topological insulator by, for instance varying the composition of the IV-VI ternary compounds, as well as highlighting the importance of the connection between bulk and surface physics in topological insulators.

\end{abstract}

\maketitle


\section{Introduction}

The Ruderman-Kittel-Kasuya-Yosida (RKKY) interaction is an indirect exchange interaction of external spins mediated by the virtual excitations of an electron gas\cite{kl14,kl15,kl16}. While the RKKY interaction is strongly suppressed in insulators due to the bulk band gap, this changes dramatically in topological insulators (TI) that feature inherent metallic surface states. These surface states are spin polarized and reflect the nontrivial topology of the bulk electronic structure, physics that will also be encoded in the RKKY interaction. However, for most topological insulators, which are rather complex materials, it is practically impossible to provide in analytical form the connection between the bulk and the surface topological states and hence to understand the surface RKKY coupling in terms of the bulk electronic structure.

As a consequence, the RKKY interaction in topological insulators has, to date, been addressed on the basis of \emph{generic} Dirac-Weyl (or, equivalently, Rashba) Hamiltonians\cite{wang15,zyu14,qin14,lit14,efi14,chei12,li12,aba11,zhu11,liu09}. These investigations have uncovered a rich RKKY physics in the topological surface state which features a wide variety of coupling interactions - Ising, Heisenberg, as well as Dzyaloshinskii-Moriya (DM). Such an approach, however, excludes the possibility of analyzing the relation between the bulk electronic structure in the surface RKKY interaction. In particular, interesting questions such as how the rich RKKY behaviour of topological insulators might be tuned by manipulating the bulk physics, e.g. by alloying, cannot be answered. In this paper we remedy this situation by establishing an explicit link between bulk band gap states and the topological surface wavefunctions for the IV-VI crystalline topological insulators\cite{vol78,vol83,vol85,pan87,saf13,tan12,tan13,tan13a,bar13,dru14,shi14,pol14}, and on this basis analyze the impact of bulk physics on the RKKY interaction.
 
We find that despite the universal form of the Dirac-Weyl topological surface state the spin structure, as probed by the RKKY interaction, is strikingly non-universal, and depends qualitatively on the crystal-field and spin-orbit parameters of the semiconductor bulk. As a specific example we show that for the well known SnTe topological insulator the form of the RKKY interaction can be strongly influenced by alloying with Pb in the bulk. 


\section{The bulk band structure model for the IV-VI compounds}

We first briefly summarize the band structure model that we will use; for further details we refer the reader to Ref.~\citeonline{vol78,vol83}. The IV-VI materials crystallize in a cubic NaCl-type lattice with 6 $p$-bands around the Fermi energy (3 conduction and 3 valence bands) that originate from the atomic $p$-shells of the group IV and group VI species.
\begin{figure}[ht]
\begin{center}
	\includegraphics[width=0.6\linewidth]{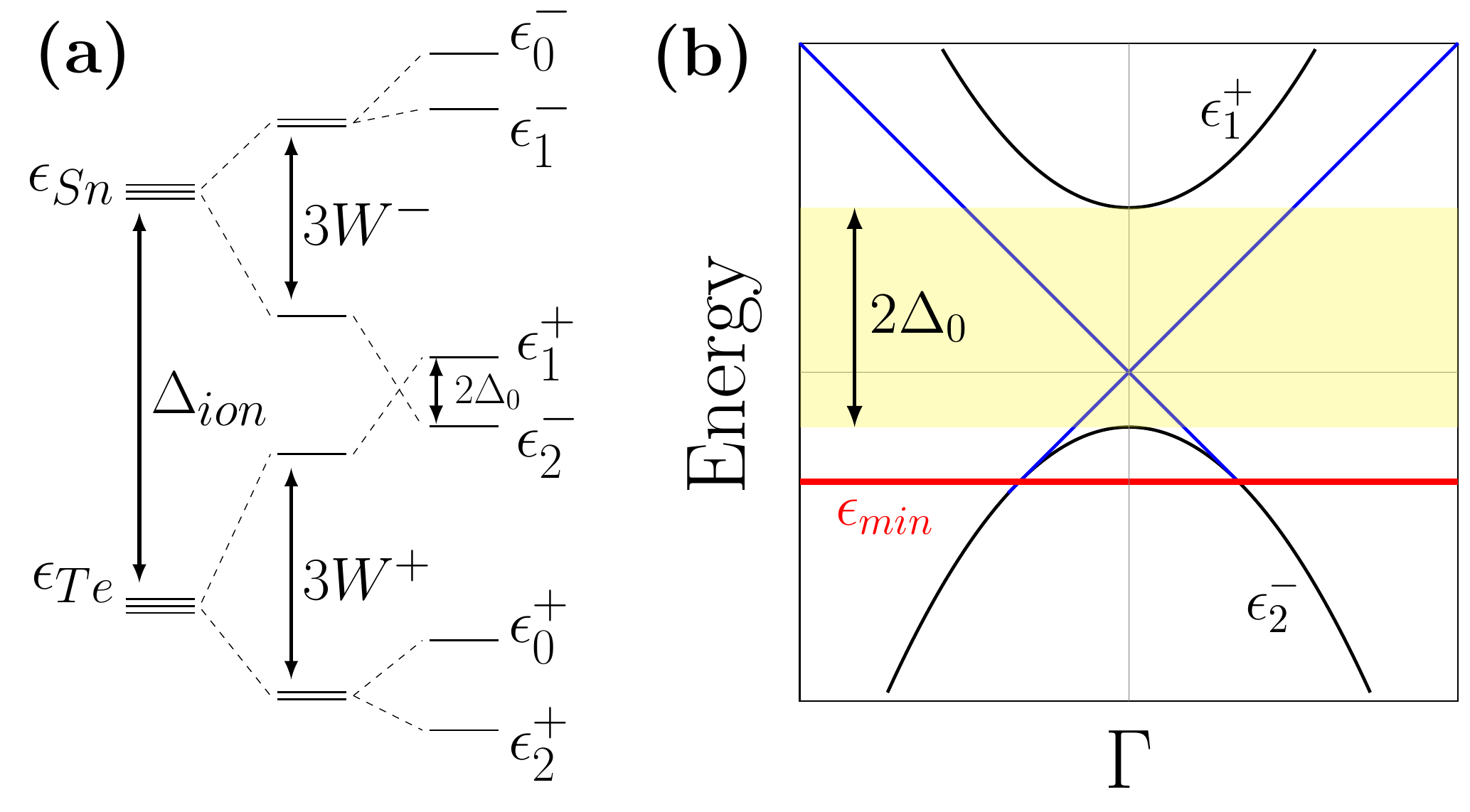}
  	\caption{Schematic illustration of the band ordering at the $L$ points (a) showing the band inversion that occurs in SnTe due to crystal field and spin orbit coupling effects. (b) low energy band structure in vicinity of the L-points exhibiting the two linear surface bands (gray) that limit at the energy \(\epsilon_{max}\) tangentially to the upper bulk band \(\epsilon_1^+\) (black).}
	\label{pic_BZ_BS}
\end{center}
\end{figure}
As shown in Ref.~\citeonline{vol78,vol83} the wave function parity at the high symmetry $L$ points provides a good quantum number and thus the Bloch function is a sum of atomic orbitals from either the group IV or the group VI species (our coordinate system well be chosen such that the group IV Bloch states have odd parity and the group VI Bloch states even parity). While the ad-mixture of the atomic $s$-states can be considerable, it does not change the \emph{form} of the effective Hamiltonian, and thus can be included by tuning the value of the model parameters to fit band structure calculated \emph{ab-initio}.
In the absence of interaction between these Bloch states, two atomic p-states would give rise to two degenerate triplets of cubically symmetric bands at the $L$ points, ordered such that the triplet of odd parity (-) group IV states is higher in energy than the triplet of even parity (+) group VI states. The lifting of the degeneracy by the crystal field matrix elements $w^\pm$ (which mix $p$-orbitals) and spin-orbit coupling $\lambda^\pm$ (which mix spin channels) reduces the band gap dramatically, as shown in the schematic illustration presented in Fig.~\ref{pic_BZ_BS}(a). 
In the case of SnTe or cubic SnSe this is sufficient to invert the ordering of the $\epsilon_1^+$ and $\epsilon_2^-$ states as compared to the vacuum, as illustrated in Fig.~\ref{pic_BZ_BS}(a), resulting in a topologically protected surface state. 

The Bloch states at the $L$-points are given by
\begin{align}
\Phi^+_{x,y,z}(\mathbf{r})=&i\sqrt{\frac{2}{N}}\sum_{\mathbf{R}} \sin\left(\mathbf{k}_L\mathbf{R}\right)p_{x,y,z}(\mathbf{r}-\mathbf{R})\label{Phi_n^+},\\
\Phi^-_{x,y,z}(\mathbf{r})=&\sqrt{\frac{2}{N}}\sum_{\mathbf{R}} \cos\left(\mathbf{k}_L\mathbf{R}\right)p_{x,y,z}(\mathbf{r}-\mathbf{R})\label{Phi_n^-}.
\end{align}
where $\bk_L = \pi/a (\pm1,\pm1,\pm1)$ is one of the eight $L$-point wave vectors and $\bR$ runs over all lattice vectors. As a result of spin and orbital angular momentum mixing, the band the band edge states at the $L$ point are given by the Kramers conjugate pairs
\begin{align}
\Phi_2^- &=-\sin\frac{\theta_-}{2}\Phi_+^{-\downarrow}+\cos\frac{\theta_-}{2}\Phi_0^{-\uparrow}\\
K\Phi_2^- &=-\sin\frac{\theta_-}{2}\Phi_-^{-\uparrow}+\cos\frac{\theta_-}{2}\Phi_0^{-\downarrow}\\
\Phi_1^+ &=\ \ \cos\frac{\theta_+}{2}\Phi_+^{+\downarrow}+\sin\frac{\theta_+}{2}\Phi_0^{+\uparrow}\\
K\Phi_1^+ &=\ \ \cos\frac{\theta_+}{2}\Phi_-^{+\uparrow}+\sin\frac{\theta_+}{2}\Phi_0^{+\downarrow}
\end{align}
where $\Phi_0^{\pm\uparrow\downarrow}=\Phi_{z'}^{\pm\uparrow\downarrow}\), \(\Phi_\pm^{\pm\uparrow\downarrow}=1/\sqrt{2}\left(\Phi_{x'}^{\pm\uparrow\downarrow}\pm i \Phi_{y'}^{\pm\uparrow\downarrow}\right)$ and where the primed coordinate system has the $z'$-axis aligned along the $(111)$ symmetry axis which forms a natural coordinate system. The \emph{spin mixing parameters} \(\theta_\pm\) are connected to the crystal field and the spin orbit interaction as follows\cite{vol83}:
\begin{align}
&\tan\theta_\pm=-\frac{2\sqrt{2}\hbar \lambda^\pm}{\hbar\lambda^\pm+3w^\pm}.
\label{SM}
\end{align}
This parameter will be subsequently shown to dramatically influence the RKKY interaction on the (111) surface of these materials.
As shown in Ref.~\citeonline{vol83}, the \(C_{3v}\) symmetry of the $L$-point dictates the following Hamiltonian
\begin{align}
H^\pm=\mp\Delta_{ion}+2w^\pm\cos\left(\frac{2\pi}{3}L_{z^\prime}\right)+\lambda^\pm\left(\mathbf{L}.\boldsymbol{\sigma}\right)^\pm
\label{eqH}
\end{align}
where the \(\pm\) index refers to the conduction band edge (group VI derived even parity state) or the valence band edge (group IV derived odd parity state) and where ${\bf L}$ is the angular momentum operator. The first term in Eq.~\ref{eqH} represents the difference in ionization energy between the group IV and group VI species; the second term encodes the mixing of the $p$-orbitals which, in the Hilbert space of the three $p$-states, is equivalent to the actions of the rotation operator by angles of \(2\pi /3\) around the $(111)$ symmetry axis; the third term is the spin-orbit coupling. The parameters \(w^\pm\) and \(\lambda^\pm\) are tabulated in Refs.~\citeonline{vol78,vol83} for all IV-VI compounds. Finally a \(\mathbf{k}.\mathbf{p}\) expansion around the $L$-point leads to a twelve-band \(\mathbf{k}.\mathbf{p}\) Hamiltonian, which reduces at low energies to a four-band Dirac-type Hamiltonian\cite{vol83}
\begin{align}
H=\begin{pmatrix}
\Delta&\hbar\left[v_\perp \boldsymbol{\tau}_\perp^\prime.\mathbf{k}'_\perp+v_\parallel \tau_z^\prime k_z^\prime\right]\\
\hbar\left[v_\perp \boldsymbol{\tau}_\perp^\prime.\mathbf{k}'_\perp+v_\parallel \tau_z^\prime k_z^\prime\right]&-\Delta
\end{pmatrix}
\label{H_D},
\end{align}
where \(\boldsymbol{\tau}_\perp^\prime\) stands for the vector of Pauli matrices $(\tau_x^\prime,\tau_y^\prime)$. In addition to the bulk band gap $\Delta$, the Hamiltonian contains \(v_\parallel\) (\(v_\perp\)), the velocity parallel (perpendicular) to the $(111)$ symmetry axis, and which are responsible for anisotropic effective masses parallel or perpendicular to this axis.


\section{The topological surface states}

To describe the low energy electron excitations on the surface we take Eq.~\ref{H_D} and replace the band gap $\Delta$ by a \emph{band gap function} \(\Delta(z)=\Delta_0f(z)\) with \(\Delta_0<0\) and $f(z)$ ($f(+\infty) = 1$ and $f(-\infty) \to -\infty$) such that $\Delta(-\infty)\Delta(\infty) < 0$, as required for a topologically non-trivial band structure\cite{vol85}. Furthermore, we introduce an energy shift function \(\varphi(z)=\varphi_0f(z)\) that models the expected overall shift in energy of the states at the surface\cite{pan87} (i.e., band bending). This yields the bulk-boundary eigenvalue problem
\begin{align}
\begin{pmatrix}
\Delta(z)& v_\parallel \tau_z p_z+v_\perp \boldsymbol{\tau}_\perp.\mathbf{p}_\perp\\
v_\parallel \tau_z p_z+v_\perp \boldsymbol{\tau}_\perp.\mathbf{p}_\perp&-\Delta(z)
\end{pmatrix}
\psi=\left(\epsilon-\varphi(z)\right)\psi,
\label{EVP_shift}
\end{align}
Note that for the experimentally observed downward band bending\cite{tas14} we require \(\varphi_0>0\).
There are four inequivalent $L$-points in the bulk with, for the (111) surface, $\bk_L=\pi/a(1,1,1)$ projecting to the $\Gamma$ point of the (111) surface Brillouin zone, and $\bk_L=\pi/a(1,1,1)$, $\bk_L=\pi/a(1,1,1)$, and $\bk_L=\pi/a(1,1,1)$ to the three inequivalent $M$-points. For reasons that will be explained below we focus on the $\Gamma$-point of the (111) surface. This choice is, in fact, already implied by Eq.~\ref{EVP_shift}: setting up the bulk-boundary problem for another crystal facet -- or a different $L$-point -- would require rotation of $k_z$ in Eq.~\ref{H_D} such that the principle axis of the ellipsoidal band manifold at the $L$-point coincides with the surface normal.

After a basis transformation Eq.~\ref{EVP_shift} takes the form of a supersymmetric Dirac equation:

\begin{align}
&h_\pm^\dagger h_\pm\zeta_\pm=\left[\epsilon^2-\hbar^2v_\perp^2k_\perp^2+\frac{\epsilon^2\varphi_0^2}{\Delta_0^2-\varphi_0^2}\right]\zeta_\pm
\label{SSE_W}
\end{align}
where $h_\pm = \left[i\sigma_zv_\parallel p_z\pm W(z)\right]$ with \(W(z)=\sqrt{\Delta_0^2-\varphi_0^2}\left[f(z)+\epsilon\varphi_0/\left(\Delta_0^2-\varphi_0^2\right)\right]\) the superpotential and where we have defined the components of two (spinor valued) parts of the transformed wavefunction  \(\zeta=\begin{pmatrix}\zeta_s, & \zeta_{-s}\end{pmatrix}\). The only normalizable solution for the ground state of the supersymmetric equation are the pure spinors $\zeta_s \propto \begin{pmatrix} 1, & 0 \end{pmatrix} g(z)$ and $\zeta_{-s} \propto \begin{pmatrix} 0, & 1 \end{pmatrix} g(z)$. Inserting \(\zeta\) into Eq. \ref{EVP_shift} leads to a differential equation in $z$ (the direction normal to the surface) that is solved by the envelope function \(g(z)\) (decaying exponentially into the vacuum and bulk). This leads to an effective Dirac-Weyl equation on the surface
\begin{align}
H_{eff}(\mathbf{k}_\perp)\chi=\epsilon \chi
\end{align}
where the effective surface Hamiltonian is given by \(H_{eff}(\mathbf{k}_\perp)=\hbar \gamma v_\perp\boldsymbol{\eta}_\perp.\mathbf{k}_\perp\) with \(\boldsymbol{\eta}\) the two-dimensional vector of Pauli matrices and \(\gamma=\sqrt{1-\varphi_0^2/\Delta_0^2}\). This equation, as is well known, describes massless Dirac fermions, and the surface spectrum consists of two linear dispersing energy bands \(\epsilon_{k_\perp l}=l\hbar \tilde{v}_\perp k_\perp\) with the velocity \(\gamma v_\perp\). The corresponding eigenstates are given by
\begin{align}
\chi_{\mathbf{k}_\perp l}= \frac{1}{\sqrt{2}}
\begin{pmatrix}
l e^{-i\frac{\phi}{2}}\\
e^{i\frac{\phi}{2}}
\end{pmatrix}
\label{chi_SnTe}
\end{align}
This solution is, however, not the physical wavefunction of the surface state. The effective surface Hamiltonian and its eigenstates are defined on a pseudospin space that is spanned by the Kramers conjugate pair $X$ and $K X$ with

\begin{equation}
X=\frac{s}{\sqrt{2}}\left[e^{-i\frac{\pi}{4}}\sqrt{1-\frac{\varphi_0}{\Delta_0}}\Phi_2^-+e^{i\frac{\pi}{4}}\sqrt{1+\frac{\varphi_0}{\Delta_0}}\Phi_1^+\right]
\end{equation} 
leading to the physical wavefunction of the surface state given by

\begin{equation}
\Psi_{\mathbf{k}_\perp l} = (c_+ X + c_- KX) e^{i\mathbf{k}_\perp.\mathbf{r}_\perp}e^{i\kappa z}g(z)
\label{yolo}
\end{equation}

The surface states are characterized by two quantum numbers, the chirality \(l\in\{\pm\}\) and the wave vector \(\mathbf{k}_\perp\) that in polar coordinates is given by \((k_\perp,\phi)\). The function $g(z)=N \exp\left(-\frac{s}{\hbar v_\parallel}\int_0^z dz\ W(z)\right)$ ensures an exponential decay away from the surface \emph{provided that the superpotential changes its sign asymptotically}. From the definition of the superpotential, it is seen that this behaviour is ensured by the function \(f(z)\) provided \(|\varphi_0|<|\Delta_0|\) and $\epsilon_{-s} < \epsilon <\epsilon_{s}$ with $\epsilon_{s}=-\varphi(-s\infty)\left(\Delta_0^2/\varphi_0^2-1\right)$. The energy shift function (\(\varphi_0(z)\)) guarantees that the two surface bands limit tangentially to the bulk band \(\epsilon_1^+(k_\perp)\) for $s = -1$, or \(\epsilon_2^-(k_\perp)\) for $s = +1$. This relation between bulk bands and the surface state is illustrated in Fig.~\ref{pic_BZ_BS}(b).

As our theory distinguishes in a clear way the physical electron spin from the pseudospin degree of freedom we can calculate expectation values of both the pseudo- as well as electron spin operators for a given surface state \(\chi_{\mathbf{k}_\perp l}\). For the pseudospin we find the usual \emph{irrotational} pseudospin density vector field $\langle\chi_{\mathbf{k}_\perp l} \mid\boldsymbol{\eta}\mid \chi_{\mathbf{k}_\perp l}\rangle=l \begin{pmatrix} \cos\phi, & \sin\phi, & 0 \end{pmatrix}$ with unit magnitude while, in contrast, the physical spin density is the well known \emph{solenoidal} field $\langle\Psi_{\mathbf{k}_\perp l} \mid \boldsymbol{\sigma}\mid\Psi_{\mathbf{k}_\perp l}\rangle=l\sigma \begin{pmatrix}-\sin\phi, & \cos\phi, & 0 \end{pmatrix}$, with a magnitude that depends explicitly on the bulk band structure parameters \(\sigma=1/2\cos\left[\left(\theta_-+\theta_+\right)/2\right]-\varphi_0/(2\Delta_0) \cos\left[\left(\theta_--\theta_+\right)/2\right]\). Note that as we have derived our surface state from a microscopic theory of the bulk crystal, the value of $\sigma$ incorporates the bulk spin-orbit coupling that is known to reduce the spin polarization of the surface state\cite{yaz10}.


\section{The RKKY Interaction}

\begin{figure}[ht]
\centering
\includegraphics[width=\linewidth]{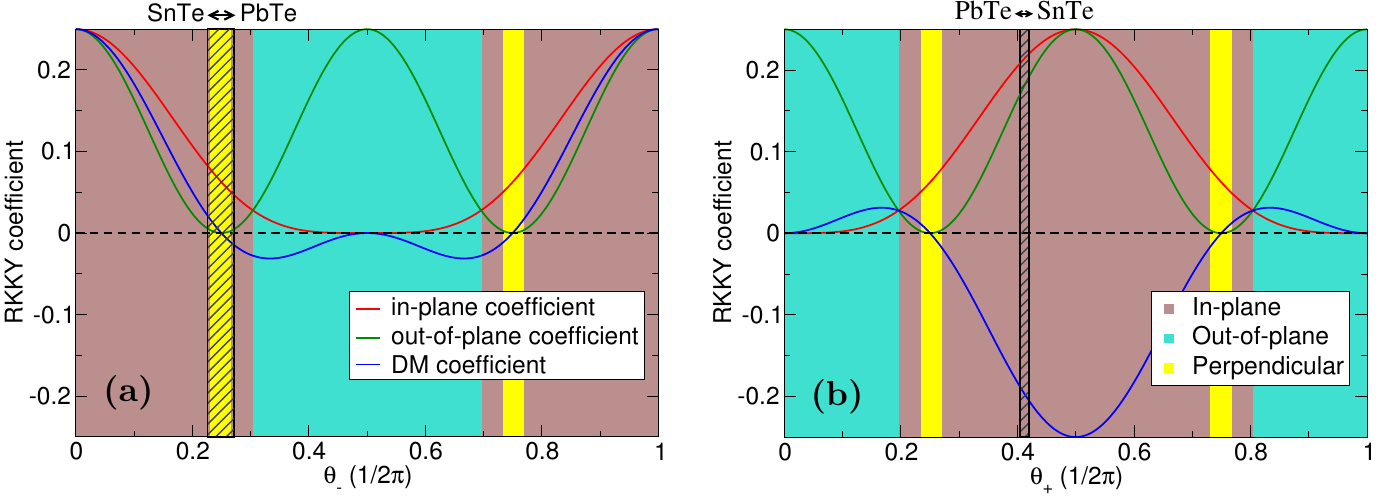}
\caption{The evolution of the strength of the RKKY coefficients \(a_\alpha\) (in-plane coupling), \(b_\alpha\) (out-of-plane coupling) and \(c_\alpha\) (Dzyaloshinskii-Moriya coupling) as a function of the spin mixing parameters \(\theta_\alpha\) of a topological insulator of the SnTe class. The parameters $\theta_\pm$ reflect the relative strengths of spin orbit coupling and crystal field in the bulk for the Sn (-) and Te (+) species (with $\theta_\pm=0$ corresponding to no spin-orbit coupling, see Eq.~\ref{SM}). The changing coupling strength of these interaction types results in three qualitatively distinct regions of the RKKY interaction, highlighted by the different background color: in the brown and in the turquoise region both spins cant with respect to a ferromagnetic or anti-ferromagnetic reference state parallel to the \(x\)- or \(z\)-axis respectively, while in the yellow region a collinear coupling parallel to the \(y\)-axis is preferred (the choice of coordinate system is such that the $x$-axis is aligned with the connection vector of the spin impurities and $z$ points out-of-plane). The values of $\theta_\pm$ that correspond to the Pb$_x$Sn$_{1-x}$Te system are indicated by the shaded area.}
\label{abc}
\end{figure}

We now come to the RKKY interaction between two magnetic moments on the surface of a IV-VI topological insulator. We will consider the $\Gamma$-point Dirac cone on the (111) surface and ignore possible contribution from the M-point cones. This is likely to be a rather good approximation as the energy separation of the $\Gamma$ and $L$-point cones ($\approx 170\,$meV), along with the limited energy range within which the Dirac cone exists ($\approx 200\,$meV), implies that for energies at which the RKKY signal from the $\Gamma$-cone is strong, that from the $M$-cone will be weak.

We consider the indirect exchange interaction to be mediated by the topological surface states, Eq.~\ref{chi_SnTe}, and assume that each impurity spin \(\mathbf{S}_{1,2}\) couples via a contact interaction to the electron spin density \(\mathbf{s}_{1,2}\)

\begin{align}
V=-\beta\left(\mathbf{S}_1.\mathbf{s}_1+\mathbf{S}_1.\mathbf{s}_2\right),
\label{eqV}
\end{align}
where the coupling strength is given by \(\beta\). As a further simplification of the model we set the energy shift function $\varphi_0 = 0$ (i.e. no band bending at the surface), an approximation expected to be good provided the Fermi energy does not approach the tangent point of the Dirac cone at which the surface state merges with the bulk spectrum.

\begin{figure*}[t!]
\centering
\includegraphics[width=0.94\linewidth]{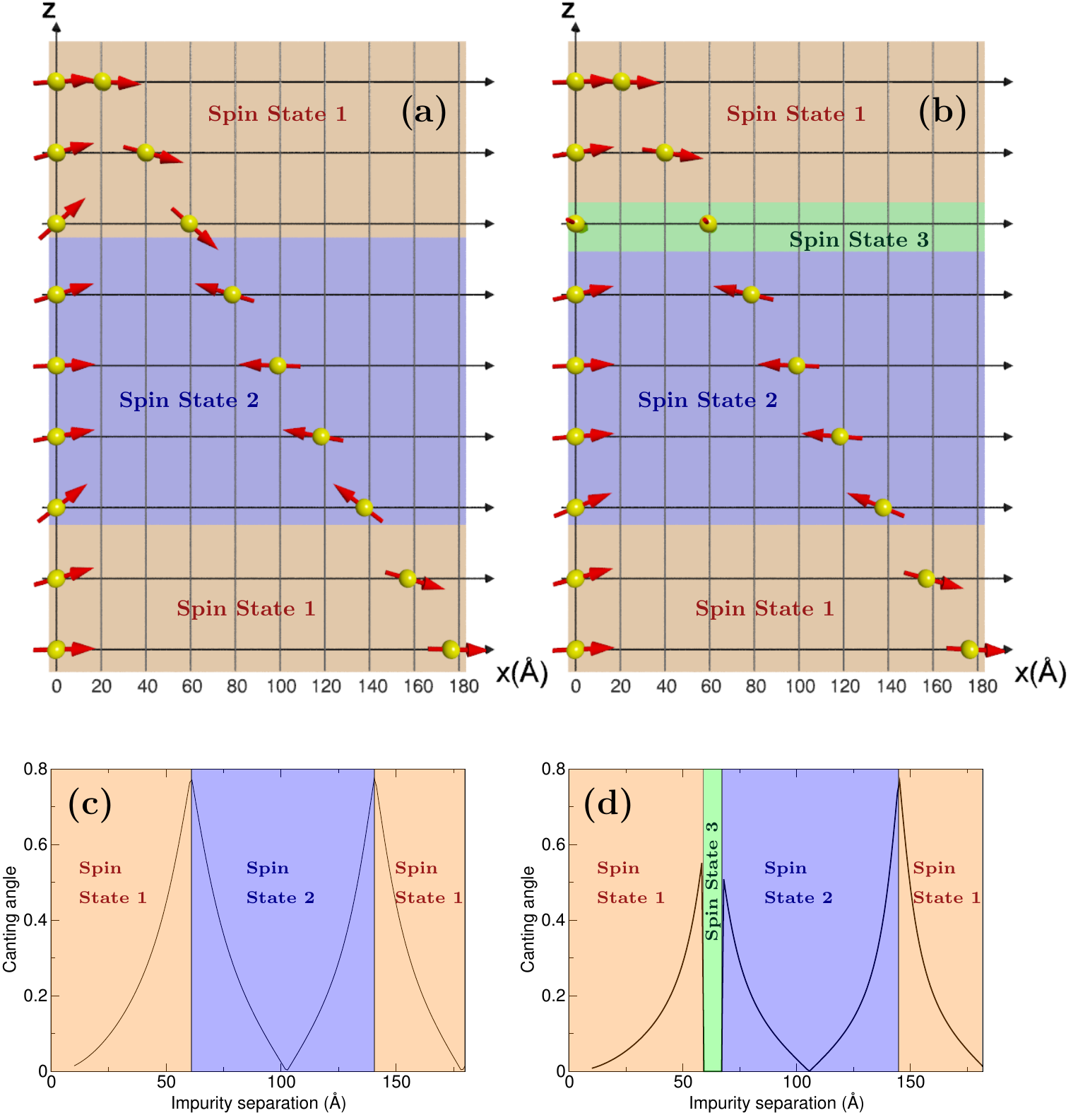}
\caption{The RKKY interaction on the tin terminated (111) surface of SnTe (panels a and c) and \(\mathrm{Pb}_{0.25}\mathrm{Sn}_{0.75}\mathrm{Te}\) (panels b and d) at a Fermi energy of \(0.1\)~eV. The SnTe crystal features two distinct spin states: canting with respect to the ferromagnetic coupling parallel to the \(x\)-axis (spin state 1) and canting with respect to the antiferromagnetic reference state parallel to the \(x\)-axis (spin state 2). The canting angle is given in panel c). By alloying the SnTe crystal with lead it is possible to open a small window where ferromagnetic coupling parallel to the \(y\)-axis is preferred (spin state 3).}
\label{cantr}
\end{figure*}

To calculate the RKKY interaction we require an understanding of the impurity scattering mechanism due to Eq.~\ref{eqV}. This is encoded in the scattering matrix $\boldsymbol{\sigma}.\mathbf{S}_1$.
For scattering at a group IV site impurity the \(\boldsymbol{\sigma}.\mathbf{S}_1\)-matrix in the pseudospin basis takes the form

\begin{align}
\left(\boldsymbol{\sigma}.\mathbf{S}_1\right)_-=
\frac{1}{2}\begin{pmatrix}
S_1^z\cos\theta_-&iS_1^-\cos^2\frac{\theta_-}{2}\\
-iS_1^+\cos^2\frac{\theta_-}{2}&-S_1^z \cos\theta_-
\end{pmatrix}
\label{sigmaS-}
\end{align}
while for scattering at a group VI site impurity it is given by

\begin{align}
\left(\boldsymbol{\sigma}.\mathbf{S}_1\right)_+=
\frac{1}{2}\begin{pmatrix}
-S_1^z\cos\theta_+&-iS_1^-\sin^2\frac{\theta_+}{2}\\
iS_1^+\sin^2\frac{\theta_+}{2}&S_1^z \cos\theta_+
\end{pmatrix}
\label{sigmaS+}
\end{align}
with \(S_1^\pm=S_1^x\pm i S_1^y\). Employing these scattering matrices with a standard RKKY calculation leads to the following interaction energy

\begin{align}
\notag E_\alpha^{int}(R_\perp)=-\frac{2\pi^2\hbar \lambda^2}{\Omega_{SBZ}^2 v_\perp}&\Bigl(\left[A_{k_F}(R_\perp)-B_{k_F}(R_\perp)\right]\left[a_\alpha S_1^xS_2^x+b_\alpha S_1^zS_2^z\right]\Bigr.
+\left[A_{k_F}(R_\perp)+B_{k_F}(R_\perp)\right]a_\alpha S_1^yS_2^y\\
&\Bigr.+\left[C_{k_F}(R_\perp)+D_{k_F}(R_\perp)\right]c_\alpha\left[S_1^xS_2^z- S_1^zS_2^x\right]\Bigl)
\label{E_int_TI}
\end{align}
where the material dependent coefficients \(a_\alpha\), \(b_\alpha\) and \(c_\alpha\) are given in Table \ref{sigma-rho} and where

\begin{align}
A_{k_F}(R_\perp)=\frac{\pi}{2}\lim_{s\to 0}\int_{k_F}^\infty dk_\perp\ k_\perp^ 2 J_0\left(k_\perp R_\perp\right)Y_0\left(k_\perp R_\perp\right)e^{-sk_\perp}\label{A1}\\
B_{k_F}(R_\perp)=\frac{\pi}{2}\lim_{s\to 0}\int_{k_F}^\infty dk_\perp\ k_\perp^ 2 J_1\left(k_\perp R_\perp\right)Y_1\left(k_\perp R_\perp\right)e^{-sk_\perp}\\
C_{k_F}(R_\perp)=\frac{\pi}{2}\lim_{s\to 0}\int_{k_F}^\infty dk_\perp\ k_\perp^ 2 J_1\left(k_\perp R_\perp\right)Y_0\left(k_\perp R_\perp\right)e^{-sk_\perp}\\
D_{k_F}(R_\perp)=\frac{\pi}{2}\lim_{s\to 0}\int_{k_F}^\infty dk_\perp\ k_\perp^ 2 J_0\left(k_\perp R_\perp\right)Y_1\left(k_\perp R_\perp\right)e^{-sk_\perp}\label{D1}
\end{align}
with \(k_\perp=E/(\hbar v_\perp)\).
Following an asymptotic expansion Eq.~\ref{E_int_TI} can be brought to the form

\begin{align}
\notag E_\alpha^{int}(R_\perp)=-\frac{\pi^2k_F\hbar\lambda^2}{\Omega_{SBZ}^2 v_\perp R_\perp^2}&\Biggl[\sin\Bigl(2k_FR_\perp\Bigl) \Bigl(a_\alpha S_1^xS_2^x+b_\alpha S_1^zS_2^z\Bigr)\Biggr.
\Biggl.-c_\alpha \cos\left(2k_FR_\perp\right)\Bigl(S_1^xS_2^z-S_1^zS_2^x\Bigr)\Biggr]\\
\notag-\frac{\pi^2\hbar\lambda^2}{4\Omega_{SBZ}^2v_\perp R_\perp^3}&\Biggl[\cos\left(2k_F R_\perp\right)\Bigl(3a_\alpha S_1^xS_2^x-2a_\alpha S_1^yS_2^y+3b_\alpha S_1^zS_2^z\Bigr)\Biggr.
\Biggl.+3c_\alpha \sin\left(2k_F R_\perp\right)\Bigl(S_1^xS_2^z-S_1^zS_2^x\Bigr)\Biggr]
\end{align}
Note that fact that in this equation (as well as in the full interaction energy Eq.~\ref{E_int_TI}) the dependence is solely through the impurity separation \(R_\perp\), and not the polar angle \(\theta\), arises from the choice of the coordinate system in which we have rotated \(x\)-axis such that it is aligned with the impurity separation vector.

\begin{table}[h!]
\begin{align*}
\arraycolsep=12pt\def\arraystretch{1.8}
\begin{array}{c|c|c}
\hline\hline
a_\alpha&b_\alpha&c_\alpha\\
\hline
\frac{1}{4}\cos^4\left(\frac{2\theta_\alpha+\pi(1+\alpha)}{4}\right)& \frac{1}{4}\cos^2\theta_\alpha& \frac{1}{4}\cos\theta_\alpha\cos^2\left(\frac{2\theta_\alpha+\pi(1+\alpha)}{4}\right)\\[0.2cm]
\hline\hline
\end{array}
\end{align*}
\caption{The coefficients for the RKKY interaction on the surface of a topological insulator of the SnTe class. The spin mixing parameter \(\theta_\alpha\) is for a range of IV-VI semiconductors given in Ref.~\citeonline{vol83}. The parameter \(\alpha\) takes the value +1 on the tellurium terminated surface and -1 on the tin terminated surface.}
\label{sigma-rho}
\end{table}

In the lowest order ($1/R_\perp^2$) there are two Ising-type terms ($S_1^xS_2^x+b_\alpha S_1^zS_2^z$) that favour a collinear ferromagnetic (FM) or antiferromagnetic (AFM) coupling, as well as a Dzyaloshinskii-Moriya (DM) term ($S_1^xS_2^z-S_1^zS_2^x$) that favours the spins to be (i) in the plane formed by the connection vector and the surface normal and (ii) have a relative angle of 90$^\circ$. Curiously, the collinear terms are $\pi/2$ out of phase with the DM term (a fact also true at order $1/R_\perp^3$). However, the most interesting aspect of the interaction energy is that each coupling term is endowed with its own material constant and, as may be seen in Fig.~\ref{abc}, these are oscillatory functions of the spin mixing angles and, hence, of the materials crystal field and spin orbit coupling parameters. The \emph{qualitative form} of the RKKY interaction in the IV-VI topological insulators is, therefore, profoundly sensitive to the details of the bulk electronic structure: there is no ``universal'' form of the RKKY interaction corresponding to, or resulting from, the universality of the Dirac-Weyl operator that describes the topological surface state.

From Fig.~\ref{abc} it can be noticed that (i) the materials parameters for the group IV terminated surface ($-$) and group VI terminated surface ($+$) are, with a sign change of the DM constant, related to each other by a phase shift of $\pi$ and (ii) for $\lambda^- = 0$, i.e. zero spin orbit coupling, the non-collinear DM coupling on the group VI terminated surface is at \emph{maximum strength} (provided, of course, that the crystal field by itself is enough to invert the band order). A strong DM interaction is usually associated with substantial spin orbit coupling; this example highlights the fact that it is the non-trivial topology of the bulk band structure, not the spin orbit coupling, that is responsible for the DM interaction.

To investigate impact of the bulk physics on the impurity spin coupling for a specific example we now consider spin impurities on the tin surface of SnTe and its alloy \(\mathrm{Pb}_{0.25}\mathrm{Sn}_{0.75}\mathrm{Te}\), illustrated in panels (a) and (b) of Fig.~\ref{cantr} respectively. On the tin terminated surface of SnTe and at an impurity separation of 20\AA~both impurity spins couple at FM and parallel to the connection vector. At higher impurity separation the two spins cant away from the connection vector leading to an increasing out-of-plane component. At an impurity separation of \(\approx80\)\AA~the canting angle reaches its maximum and the configuration flips into an AFM coupling, following which the canting angle then begins to decrease. In the further course of the evolution with increasing separation the two spins oscillate between these two types of spin states with successively increasing and decreasing canting angle, as shown in \ref{cantr}(c). The period of oscillation is given by the Fermi energy which is here set to \(0.1\)~eV. The corresponding behaviour on the tin terminated surface of \(\mathrm{Pb}_{0.25}\mathrm{Sn}_{0.75}\mathrm{Te}\) is shown in Fig. \ref{cantr}(b): in large parts the interaction resembles that on the surface of SnTe but in small window from \(59\)\AA~to \(67\)\AA~a coupling in the direction parallel to the \(y\)-axis can be observed. As shown in Fig.~\ref{cantr}(d) the canting angle is in that region rather small and thus the coupling can considered to be, as a good approximation, ferromagnetic.

\section{Discussion and conclusions}

We have employed the RKKY interaction to probe the coupling between the spin structure of the surface state and that of the bulk insulator wavefunctions in the IV-VI semiconductor crystalline topological insulators. For the (111) surface we have derived the Dirac-Weyl surface state from a fully microscopic description of the bulk insulator, finding in this way an explicit connection between the surface and bulk wavefunctions. This, in turn, provides a ``fully electronic'' route from the atomic degrees of freedom of the bulk to the emergent pseudospin structure of the surface state and allows one to probe the impact of bulk physics on the surface RKKY interaction.

We find that the nature of the topological surface state, as probed by the RKKY interaction, is coupled very strongly to the bulk physics: the balance of crystal field and spin-orbit coupling in the bulk insulator determines even the qualitative form of the RKKY interaction. For the group IV terminated surface, when the crystal field dominates spin-orbit in the bulk the RKKY interaction favours an out-of-plane configuration, while when the the reverse is true an in-plane configuration is favoured. On the group VI terminated surface the situation is reversed. The equilibrium spin configuration is generally non-collinear, with collinear spins found only when the spin-orbit coupling in the bulk is switched off. In short, while the Dirac-Weyl surface state is ``universal'' in the sense that the surface state spectrum is always, at low energies, a conical intersection, the surface wavefunction is not and depends strongly on details of the bulk electronic structure, a fact reflected in the rich dependence of the type of RKKY interaction on the microscopic parameters of the bulk band structure.

\bibliographystyle{apsrev}
\bibliography{SnTe111_paper}

\end{document}